\documentclass[12pt]{article}

\pdfoutput = 1

\usepackage{amsmath}
\usepackage{graphicx}
\usepackage{subcaption}
\usepackage{titlesec}
\usepackage{bigints}
\usepackage{xcolor}
\usepackage{hyperref}

\titlelabel{\thetitle.\quad}
\numberwithin{equation}{section}

\begin{document}

\title{{\bf{\Large Anisotropic scalar-metric quantum cosmology and unitarity}}}
\author{
	{\bf {\normalsize Saumya Ghosh}$^{a}
	$\thanks{sgsgsaumya@gmail.com, sg14ip041@iiserkol.ac.in}}
	{\bf {\normalsize , Sunandan Gangopadhyay}$^{b}
	$\thanks{sunandan.gangopadhyay@gmail.com, sunandan.gangopadhyay@bose.res.in}}
	{\bf {\normalsize , Prasanta K. Panigrahi}}$^{a}
	$\thanks{pprasanta@iiserkol.ac.in}\\
	$^{a}${\normalsize\textit{Department of Physical Sciences,}}\\
	{\normalsize\textit{Indian Institute of Science Education and Research Kolkata,}}\\
	{\normalsize\textit{Mohanpur, Nadia, West Bengal, 741 246, India.}}\\
	$^{b}${\normalsize\textit{Department of Theoretical Sciences,}}\\
	{\normalsize\textit{S.N. Bose National Centre for Basic Sciences,}}\\
	{\normalsize\textit{Block - JD, Sector - III, Salt Lake, Kolkata - 700106, India.}}\\
       }
\date{}

\maketitle

\begin{abstract}
\noindent In this article we perform the Wheeler-DeWitt quantization for Bianchi type $I$ anisotropic cosmological model in the presence of a scalar field minimally coupled to the Einstein-Hilbert gravity theory. We also consider the cosmological (perfect) fluid to construct the matter sector of the model whose dynamics plays the role of time. After obtaining the Wheeler-DeWitt equation from the Hamiltonian formalism, we then define the self-adjointness relations properly. Doing that we proceed to get a solution for the Wheeler-DeWitt equation and construct a well behaved wave function for the universe. The unitary evolution of the wave function in the presence of the scalar field is finally restored by an appropriate choice of weight factor in defining the norm of the wave packet. It is then concluded that the Big-Bang singularity can be removed in the context of quantum cosmology. 
\end{abstract}
\vskip 1cm

\section{Introduction}
Quantum cosmology which is the canonical quantum theory of cosmology applied to the universe originated in the works of Wheeler and DeWitt \cite{J. A. Wheeler, B. S. DeWitt}. The scope of quantum cosmology is two fold. The first is to understand the birth of the universe. In particular the Hartle-Hawking no-boundary proposal \cite{Hartle et. al.} and Vilenkin's tunneling proposal \cite{A. Vilenkin} are the direct consequences of quantum cosmology. The second one is the hope that a quantum theory of the universe would lead to the removal of the big-bang singularity.  
A step toward understanding both these fundamental questions have been taken in the following way.
It is believed that the behaviour of the universe at very early stages of evolution is governed by the Wheeler-DeWitt (WD) equation. However, it has also been realized that there are many conceptual and technical problems in this description. One of the difficulties involved is the identification of a time parameter \cite{kuchar}-\cite{rov}. Problem related to the interpretation of the wave function is another issue that remains at the heart of quantum cosmology 
\cite{coleman},\cite{Pinto-Neto et. al.}. The problem of time gets resolved by the introduction of the matter sector of the cosmological model. Following the scheme developed by Schutz \cite{B. Schutz 01, B. Schutz 02}, the matter field variable can be made to play the role of time in the WD equation. This was first carried out in \cite{Lapchinskii et. al}. 
In this scheme one usually considers the cosmological matter to obey a perfect fluid equation of state \cite{Alvarenga et. al.}-\cite{Pal et. al. 02}. The role of time is played by the monotonic evolution of the fluid density. However, the Chaplygin gas model has also been used as the matter sector in many studies \cite{Pedram et. al.}-\cite{Saumya et. al.}. The reason for using the Chaplygin gas model is due to its ability to describe the accelerating stage of the universe \cite{Kamenshchik et. al., Bento et. al.}. Further, the quantization procedure has been carried out for both the isotropic \cite{Alvarenga et. al. 02}-\cite{Vakili et. al.} and anisotropic cosmological models \cite{Alvarenga et. al., Pal et. al. 01, Pal et. al. 02}.
In the context of anisotropic cosmological models, it has been realized that the quantization of these models have a problem involving the nonunitary evolution of the system which in turn implies a non-conservation of probability. To be more specific, one finds that the norm of the wave function is time dependent in the Bianchi $I$, $V$ and $IX$ models discussed in \cite{Alvarenga et. al.} and \cite{Majumdar et. al.} respectively.
However, in \cite{Pal et. al. 01} it was shown that this nonunitarity could be removed and is not a pathology inherent to anisotropic cosmological models.

In this paper, we investigate the quantum cosmology of a scalar field coupled to an anisotropic Bianchi $I$ spacetime in the presence of a perfect fluid. The motivation of carrying out this analysis is that such an analysis has been carried out in \cite{Saumya et. al.} in the case of a flat FRW spacetime which is an isotropic model. However, one may argue that it is not certain whether the observed universe is isotropic at very early stages of its evolution. Therefore, carrying out a similar analysis in an anisotropic cosmological scenario seems to be essential. 
The reason for introducing the scalar field in the model (as was done in our previous paper 
\cite{Saumya et. al.}) is to explore its effect on the evolution of the universe. Such a curiosity is worthwhile in its own right since it is quite plausible that the evolution of the universe
at very early stages could get affected by such fields.
Once again we apply the Schutz formalism along with the canonical approach \cite{Lapchinskii et. al} to obtain the super Hamiltonian. The quantization of this Hamiltonian leads to the WD equation. We then solve this equation for the case of the stiff fluid to obtain the quantum cosmolgy wave functions of the universe. With these solutions, we then construct the wave packet. The important ingredient in the construction of the wave packet is the appropriate choice of the weight factor which makes the norm of the wave packet independent of time.




The paper is organized as follows. In section 2 we discuss the basic set up for the canonical quantization of the gravity model and also the Schutz formalism for dealing with the matter sector.
In section 3 we have carried out the quantization of the anisotropic metric-scalar theory of gravity in the presence of perfect fluid as the cosmological matter. We conclude in section 4. 

\section{Basic formalism for quantization}
Let us start by writing down the relevant action for gravity together with a scalar field minimally coupled to it and the matter sector consisting of a perfect fluid. The form of such an action is as follows
\begin{eqnarray}
\label{2.1}
\mathcal{A}&=&\int_{M}d^4x\sqrt{-g}\left[ R-F(\phi)g^{\mu\nu}\partial_{\mu}\phi\partial_{\nu}\phi\right]+2\int_{\partial M}d^3x\sqrt{h}h_{ij}K^{ij}+\int_{M}d^4x\sqrt{-g}P\\ \nonumber
&\equiv& \mathcal{A}_g+\mathcal{A}_m
\end{eqnarray}
where $K^{ij}$ is the extrinsic curvature tensor, $h_{ij}$ is the induced metric on the time-like hypersurface and $F(\phi)$ is an arbitrary function of the scalar field $\phi$. The last term describes the matter field action $\mathcal{A}_m$, $P$ being the pressure of the cosmic fluid modeled as a perfect fluid in the subsequent discussion. The equation of state of the fluid is $P=\omega\rho$, $\rho$ being the density of the fluid. We shall now use the formalism developed in \cite{B. Schutz 01} to deal with the matter sector of the theory. In this formalism, one can cast the fluid's four velocity vector in terms of four potentials $h, \epsilon, \theta$ and $S$ in the following way
\begin{eqnarray}
\label{2.2}
u_\nu=\frac{1}{h}(\partial_{\nu}\epsilon+\theta \partial_{\nu}S)
\end{eqnarray}
where $h$ is the specific enthalpy and $S$ is the specific entropy. The other two potentials $\theta$ and $\epsilon$ are irrelevant physically. The fluid four velocity is normalized as
\begin{eqnarray}
\label{2.3}
u_\nu u^\nu=1~.
\end{eqnarray} 
We shall now make use of the basic thermodynamic relations given in \cite{B. Schutz 02}. These are
\begin{subequations}
\begin{eqnarray}
\label{2.4a}
\rho & = & \rho_0(1+\Pi) \\
\label{2.4b}
h & = & 1+\Pi+P/\rho_0 \\
\label{2.4c}
\tau dS & = & d\Pi+Pd(1/\rho_0)
\end{eqnarray}
\end{subequations}
where $\rho$ is the total mass energy density of the fluid mentioned earlier, $\tau$ is the temperature, $\rho_0$ is the rest mass density and $\Pi$ is the specific internal energy of the fluid. With the help of the fluid equation of state and eq.(\ref{2.4a}), eq.(\ref{2.4c}) can be rewritten as 
\begin{equation}
\label{2.5}
\tau dS=(1+\Pi)d[\ln(1+\Pi)-\omega \ln\rho_0]~.
\end{equation}
Hence $\tau$ and $S$ can be identified to be
\begin{eqnarray}
\label{2.6}
\tau=1+\Pi~,~S=\ln(1+\Pi)-\omega \ln\rho_0~.
\end{eqnarray}
These relations together with the thermodynamic relations can be used now to obtain an expression for the pressure $P$ in terms of the specific enthalpy $h$ and specific entropy $S$. This reads
\begin{eqnarray}
\label{2.7}
P=\frac{\omega}{(1+\omega)^{1+1/\omega}}h^{1+1/\omega}e^{-S/\omega}~.
\end{eqnarray}
We will use this expression in what follows subsequently. 

\noindent We now consider the gravity sector. Here we shall consider a time dependent anisotropic cosmological model. We take up the Bianchi $I$ metric which gives an anisotropic evolution of the universe. The metric for this model is given by \cite{Pal et. al. 01} 
\begin{eqnarray}
\label{2.8}
ds^2=N^2(t)dt^2-A^2(t)dx^2-B^2(t)dy^2-C^2(t)dz^2
\end{eqnarray}
where $N(t)$ is called the lapse function and $A(t), B(t), C(t)$ are three functions of the cosmic time $t$. The Ricci scalar for this metric is given by
\begin{equation}
\label{2.9}
R=\frac{-2\left[NA\dot{B}\dot{C} + B\left(N\dot{A}\dot{C}+NA\ddot{C}-\dot{N}A\dot{C}\right)+C\left\{N\left(B\ddot{A}+\dot{A}\dot{B}+A\ddot{B}\right)- \dot{N}\left(B\dot{A}+A\dot{B}\right)\right\}\right]}{N^3ABC}
\end{equation}
where the dots denote derivative with respect to time $t$. Substituting this into the gravity part of the action (\ref{2.1}) and throwing away a total time derivative term together with considering the scalar field to be time dependent only, the gravity sector of the action along with the scalar field can be written down upto a constant volume factor as
\begin{equation}
\label{2.10}
\mathcal{A}_g=\int dt\left[-\frac{2}{N}(\dot{A}\dot{B}C+\dot{B}\dot{C}A+\dot{C}\dot{A}B)-\frac{F(\phi)ABC}{N}\dot{\phi}^2\right]~.
\end{equation}
The expression inside the square bracket can be easily identified as the Lagrangian for the gravity part. We now proceed to find out the Hamiltonian for the gravity sector. In order to do so we make the following transformations \cite{Alvarenga et. al.}
\begin{eqnarray}
\label{2.11}
A(t)=e^{Z_0+Z_++\sqrt{3}Z_-} \\ \nonumber
B(t)=e^{Z_0+Z_+-\sqrt{3}Z_-} \\ 
C(t)=e^{Z_0-2Z_+} \nonumber
\end{eqnarray}
where $Z_0(t), Z_+(t), Z_-(t)$ are the new variables that we shall work with instead of $A(t), B(t), C(t)$. In terms of these new variables, the Lagrangian takes the form 	
\begin{equation}
\label{2.12}
L_g=-\frac{6e^{3Z_0}}{N}(\dot{Z}_0^2-\dot{Z}_+^2-\dot{Z}_-^2)-\frac{F(\phi)e^{3Z_0}}{N}\dot{\phi}^2~.
\end{equation}
The Hamiltonian for the gravity sector therefore reads 
\begin{eqnarray}
\label{2.13}
H_g=-\frac{1}{24}Ne^{-3Z_0}(p_0^2-p_+^2-p_-^2)-\frac{1}{4F(\phi)}Ne^{-3Z_0}p_\phi^2
\end{eqnarray}
where $p_0, p_+, p_-$ and $p_\phi$ are the canonical momenta conjugate to $Z_0, Z_+, Z_-$ and $\phi$ respectively. The lapse function $N$ can be considered as a Lagrange multiplier in this expression for $H_g$.  

\noindent Once again we look at the matter sector of the theory. In case of an observer comoving with the cosmic fluid, it is the time evolution of the universe that only matters. Hence with respect to a comoving observer, the fluid four velocity vector takes the form $u_\nu=(N, 0, 0, 0)$. Using eq.(s)(\ref{2.2}), (\ref{2.3}), we obtain
\begin{equation}
\label{2.12a}
h=\frac{\dot{\epsilon}+\theta \dot{S}}{N}~.
\end{equation}
Substituting $h$ in eq.(\ref{2.7}) leads to the form of the matter sector of the action (\ref{2.1}) which upto a spatial volume factor reads 
\begin{eqnarray}
\label{2.14}
\mathcal{A}_m=\int dt \left[N(t)^{-1/\omega}e^{3Z_0}\frac{\omega}{(1+\omega)^{1+1/\omega}}(\dot{\epsilon}+\theta\dot{S})^{1+1/\omega}e^{-S/\omega}\right]~.
\end{eqnarray}
The Hamiltonian for the matter sector can be obtained from the matter Lagrangian following from the above action. This reads
\begin{equation}
H_m=Ne^{-\omega Z_0} p_\epsilon^{\omega+1}e^S
\end{equation}
where $p_\epsilon=\frac{\partial L_m}{\partial \dot{\epsilon}}, p_S=\frac{\partial L_m}{\partial \dot{S}}$. As mentioned in \cite{Lapchinskii et. al}, one can recast the Hamiltonian for the matter sector in a more tractable form. For that one needs the canonical transformations
\begin{subequations}
\begin{eqnarray}
\label{2.14b}
T & = & p_Se^{-S}p_\epsilon^{-(\omega+1)} \\
p_T & = & p_\epsilon^{\omega+1}e^S \\
\bar{\epsilon} & = & \epsilon-(\omega+1)\frac{p_S}{p_\epsilon} \\
\bar{p_\epsilon} & = & p_\epsilon~.
\end{eqnarray}
\end{subequations}
The Hamiltonian for the matter sector now becomes
\begin{equation}
\label{2.15}
H_m=Ne^{-3Z_0}e^{3(1-\omega)Z_0}p_T
\end{equation}
where $p_T$ is the canonical momentum conjugate to the variable $T$ which can be considered as the new cosmic time. 
Combining this with eq.(\ref{2.13}), the Hamiltonian for the full theory takes the form
\begin{eqnarray}
\label{2.16}
H\equiv H_g+H_m=Ne^{-3Z_0}\left[-\frac{1}{24}(p_0^2-p_+^2-p_-^2)-\frac{1}{4F(\phi)}p_\phi^2+e^{3(1-\omega)Z_0}p_T\right]~.
\end{eqnarray}
Now varying the full action of the theory with respect to the lapse function $N$ gives the Hamiltonian constraint
\begin{eqnarray}
\label{2.17}
\mathcal{H}=\frac{1}{N}H=0~.
\end{eqnarray}  
It also should be noted that the gauge choice $N=e^{3\omega Z_0}$ makes the new canonical variables ($T,p_T$) decouple from the gravity sector. Also from the classical point of view, the variable $T$ has the same orientation and signature as that of the cosmic time $t$. So one can easily take $T$ as the new time for the system.  So the new set of spacetime coordinates are ($Z_0, Z_+, Z_-, T$).

\section{Quantization of the model}
In this section we switch from classical to quantum physics by writing down the Wheeler-DeWitt (WD) equation for the Hamiltonian (\ref{2.16}) written down in the previous section. To get the WD equation, we first replace the momenta appearing in the Hamiltonian (\ref{2.16}) by their quantum mechanical operator representations, namely, $p_0=-i\frac{\partial}{\partial Z_0},~ p_+=-i\frac{\partial}{\partial Z_+},~ p_-=-i\frac{\partial}{\partial Z_-},~ p_\phi= -i\frac{\partial}{\partial \phi}$ and $p_T=-i\frac{\partial}{\partial T}$ respectively (setting $\hbar=1$). The WD equation then reads
\begin{equation}
\label{3.1}
\hat{H}\Psi(Z_0, Z_+, Z_-, T)=0
\end{equation}
where
\begin{equation}
\label{3.2}
\hat{H}=\left[\frac{\partial^2}{\partial Z_0^2}-\frac{\partial^2}{\partial Z_+^2}-\frac{\partial^2}{\partial Z_-^2}+\frac{1}{4F(\phi)}\frac{\partial^2}{\partial \phi^2}-24ie^{3(1-\omega)Z_0}\frac{\partial}{\partial T}\right]~.
\end{equation}   
We shall now consider a stiff fluid for which $\omega=1$. The WD equation then reduces to
\begin{equation}
\label{3.3}
\frac{\partial^2 \Psi}{\partial Z_0^2}-\frac{\partial^2 \Psi}{\partial Z_+^2}-\frac{\partial^2 \Psi}{\partial Z_-^2}+\frac{1}{4F(\phi)}\frac{\partial^2 \Psi}{\partial \phi^2}=24i\frac{\partial \Psi}{\partial T}~.
\end{equation}
We now make the following ansatz to solve the above equation  
\begin{equation}
\label{3.4}
\Psi(Z, \phi, T)=e^{-iET}\Phi(Z, \phi)~, Z\equiv(Z_0, Z_+, Z_-)~.
\end{equation}
This yields
\begin{equation}
\label{3.5}
\hat{\mathcal{H}}\Phi=24E\Phi
\end{equation}
where
\begin{equation}
\label{3.5a}
\hat{\mathcal{H}}=\frac{\partial^2}{\partial Z_0^2}-\frac{\partial^2}{\partial Z_+^2}-\frac{\partial^2}{\partial Z_-^2}+\frac{1}{4F(\phi)}\frac{\partial^2}{\partial \phi^2}~.
\end{equation}
To find the solution of this equation, we apply the separation of variables method once again 
\begin{equation}
\label{3.6}
\Phi(Z, \phi)=\xi(Z)\eta(\phi)~.
\end{equation}
Substituting this in eq.(\ref{3.5}) gives two differential equations in $\xi$ and $\eta$ 
\begin{eqnarray}
\label{3.7}
\frac{\partial^2 \xi(Z)}{\partial Z_0^2}-\frac{\partial^2 \xi(Z)}{\partial Z_+^2}-\frac{\partial^2 \xi(Z)}{\partial Z_-^2}&=&(\kappa^2+24E)\xi(Z)
\end{eqnarray}
\begin{equation}
\label{3.7a}
\frac{d^2 \eta(\phi)}{d \phi^2}+4\kappa^2F(\phi)\eta(\phi)=0 
\end{equation}
where $\kappa^2$ is the separation constant. Using the method of separation of variables in the first equation
\begin{eqnarray}
\xi(Z)=\xi_0(Z_0)\xi_+(Z_+)\xi_-(Z_-) 
\end{eqnarray}
leads to the following equations
\begin{equation}
\label{3.9a}
\frac{d^2 \xi_+(Z_+)}{dZ_+^2}+K_+^2\xi_+(Z_+)=0 
\end{equation}
\begin{equation}
\label{3.9b}
\frac{d^2 \xi_-(Z_-)}{dZ_-^2}+K_-^2\xi_-(Z_-)=0 
\end{equation}
\begin{equation}
\label{3.9c}
\frac{d^2 \xi_0(Z_0)}{d Z_0^2}+(K_+^2+K_-^2-\kappa^2-24E)\xi_0(Z_0)=0~.
\end{equation}
The solution of $\xi(Z)$ therefore becomes
\begin{eqnarray}
\label{3.10}
\xi(Z)=C_0C_+C_-e^{-iK_+Z_+}e^{-iK_-Z_-}e^{-iK_0Z_0} 
\end{eqnarray}
where
\begin{eqnarray}
\label{3.10a}
K_0^2&=&K_+^2+K_-^2-\kappa^2-24E
\end{eqnarray}
 and $C_0, C_+, C_-$ are integration constants. 
 
\noindent Now to get a solution for eq.(\ref{3.7a}), we assume $F(\phi)=\frac{\lambda}{4} \phi^m, (m\neq -2, \lambda > 0)$. With this we arrive at the solution for $\eta(\phi)$ to be
\begin{eqnarray}
\label{3.11}
\eta(\phi)=c_1 (m+2)^{-\frac{1}{m+2}}\lambda^{\frac{1}{2(m+2)}} \kappa ^{\frac{1}{m+2}} \phi^{\frac{1}{2}} \Gamma \left(1-\frac{1}{m+2}\right)
J_{-\frac{1}{m+2}}\left(\frac{2\sqrt{\lambda} \phi^{\frac{m+2}{2}} \kappa }{m+2}\right)\\ \nonumber
+c_2(m+2)^{-\frac{1}{m+2}}\lambda^{\frac{1}{2(m+2)}} \kappa ^{\frac{1}{m+2}} \phi^{\frac{1}{2}} \Gamma\left(1+\frac{1}{m+2}\right) J_{\frac{1}{m+2}}\left(\frac{2\sqrt{\lambda} \phi^{\frac{m+2}{2}} \kappa}{m+2}\right)
\end{eqnarray}
where $c_1$ and $c_2$ are constants of integration. 

\noindent To construct a well behaved wave function, we now need to discuss about the appropriate boundary conditions. To do that, we take note of the fact that for the operator $\hat{\mathcal{H}}$ to be a self-adjoint operator, one should define the inner product between any two wave functions $\Phi_1$ and $\Phi_2$ in the following way \cite{Saumya et. al.}
\begin{eqnarray}
\label{A1}
(\Phi_1,\Phi_2)=\int \Phi_1^*(Z, \phi)F(\phi)\Phi_2(Z, \phi)~dZd\phi~.
\end{eqnarray}
Since we require $\mathcal{\hat{H}}$ to be self-adjoint, we must have
\begin{equation}
\label{A3}
(\mathcal{\hat{H}}\Phi_1,\Phi_2)=(\Phi_1,\mathcal{\hat{H}}\Phi_2)~.
\end{equation}
The left hand side of the above relation reads (using the definition of the inner product)
\begin{eqnarray}
\label{A4}
(\mathcal{\hat{H}}\Phi_1,\Phi_2)&=&\int (\mathcal{\hat{H}}\Phi_1)^*F(\phi)\Phi_2~dZd\phi \\ 
\label{A4.1}
&=&\bigintsss \left[\frac{\partial^2 \Phi_1^*}{\partial Z_0^2}-\frac{\partial^2 \Phi_1^*}{\partial Z_+^2}-\frac{\partial^2 \Phi_1^*}{\partial Z_-^2}+\frac{1}{4F(\phi)}\frac{\partial^2 \Phi_1^*}{\partial \phi^2}\right] F(\phi) \Phi_2~d\phi dZ~.
\end{eqnarray}
We now concentrate on the first term in the above expression which reads
\begin{eqnarray}
\label{A5}
&\qquad&\bigintsss\left(\frac{\partial^2 \Phi_1^*}{\partial Z_0^2}F(\phi) \Phi_2\right)~d\phi dZ_0dZ_+dZ_-\nonumber \\ 
&=&\bigintsss \bigg\{\left[\frac{\partial \Phi_1^*}{\partial Z_0} \Phi_2\right]_{Z_0=-\infty}^\infty-\int \frac{\partial \Phi_1^*}{\partial Z_0}\frac{\partial \Phi_2}{\partial Z_0}~dZ_0\bigg\}~F(\phi)~d\phi dZ_+dZ_-\nonumber \\
&=&-\bigintsss \frac{\partial \Phi_1^*}{\partial Z_0}\frac{\partial \Phi_2}{\partial Z_0}~dZ_0~F(\phi)~d\phi dZ_+dZ_- \nonumber \\
&=&-\bigintsss\bigg\{\left[\Phi_1^* \frac{\partial\Phi_2}{\partial Z_0}\right]_{Z_0=-\infty}^\infty-\int\Phi_1^*\frac{\partial^2 \Phi_2}{\partial Z_0^2}~dZ_0\bigg\}~F(\phi)~d\phi dZ_+dZ_-\nonumber \\
&=&\bigintsss\left(\frac{\partial^2\Phi_1^*}{\partial Z_0^2}F(\phi) \Phi_2\right)~d\phi dZ_0dZ_+dZ_-\nonumber \\
&=&\bigintsss\left(\Phi_1^*F(\phi)\frac{\partial^2 \Phi_2}{\partial Z_0^2}\right)~d\phi dZ_0dZ_+dZ_-
\end{eqnarray}
where we have integrated by parts in the second line over the integration variable $Z_0$ and imposed the boundary condition $\Phi_2=0$ at $Z_0=\pm\infty$. We have once again integrated by parts in the fourth line over the variable $Z_0$ and imposed the boundary condition $\frac{\partial\Phi_2}{\partial Z_0}=0$ at $Z_0=\pm\infty$.
This procedure can be carried out for the second and third terms also in eq.(\ref{A4.1}) with the boundary conditions $\Phi_2=0$ and $\frac{\partial\Phi_2}{\partial Z_+}=0$ at $Z_+=\pm\infty$  and $\Phi_2=0$ and $\frac{\partial\Phi_2}{\partial Z_-}=0$ at $Z_-=\pm\infty$. For the last term in eq.(\ref{A4.1}), a similar approach can be followed but the boundary conditions that needs to be imposed are $\Phi_2=0$ and $\frac{\partial\Phi_2}{\partial\phi}=0$ at $\phi=0,~\infty$. With these conditions the relation (\ref{A3}) holds ensuring the operator (\ref{3.5a}) to be self-adjoint.

\noindent With the above boundary conditions in hand, we find that we should keep only the second term in $\eta(\phi)$. Therefore
\begin{equation}
\label{3.12}
\eta(\phi)=C_{m,\lambda} \kappa ^{\frac{1}{m+2}} \phi^{\frac{1}{2}} J_{\frac{1}{m+2}}\left(\frac{2\sqrt{\lambda} \phi^{\frac{m+2}{2}} \kappa}{m+2}\right)
\end{equation}
where
\begin{eqnarray}
\label{3.13}
C_{m,\lambda}=c_2(m+2)^{-\frac{1}{m+2}}\lambda^{\frac{1}{2(m+2)}} \Gamma\left(1+\frac{1}{m+2}\right)~.
\end{eqnarray}
The total wave function now becomes
\begin{eqnarray}
\label{3.14}
\Psi(Z,\phi,T)=C_0C_+C_-C_{m,\lambda} \kappa ^{\frac{1}{m+2}} \phi^{\frac{1}{2}} e^{-iK_+Z_+}e^{-iK_-Z_-}e^{-iK_0Z_0}e^{-iET}J_{\frac{1}{m+2}}\left(\frac{2\sqrt{\lambda} \phi^{\frac{m+2}{2}} \kappa}{m+2}\right)~.
\end{eqnarray}
We now proceed to construct a wave packet using the superposition principle in the following way
\begin{eqnarray}
\label{3.15}
\Psi_{wp}=\int \kappa^{\frac{1}{2}-\gamma}e^{-(K_0^2 + K_+^2 + K_-^2+\kappa^2)}\Psi(Z,\phi,T)~d\kappa dK_0 dK_+ dK_-
\end{eqnarray}
where the limits of integration for $K$'s are from $-\infty$ to $+\infty$ and for $\kappa$ from $0$ to $\infty$. The important ingredient in the above construction is the inclusion of the exponential weight factor inside the integral.
With this wave packet, we calculate its norm. This reads
\begin{eqnarray}
\label{3.16}
||\Psi_{wp}||=\int dZ d\phi \int \kappa^{\frac{1}{2}-\gamma}e^{-(K_0^2 + K_+^2 + K_-^2+\kappa^2)}\Psi(Z,\phi,T)~d\kappa dK_0 dK_+ dK_- \\ \nonumber
\times ~ F(\phi)\int \kappa'^{\frac{1}{2}-\gamma}e^{-(K_0^{'2} + K_+^{'2} + K_-^{'2}+{\kappa'}^2)}\Psi^{'*}(Z,\phi,T)~d\kappa' dK'_0 dK'_+ dK'_-~. 
\end{eqnarray}
We shall first perform the integration over the variable $dZ$. This turns out to be
\begin{gather}
\label{3.17}
\int dZd\phi~ \Psi(Z,\phi,T) F(\phi) \Psi^{'*}(Z,\phi,T) \\ \nonumber
=(C_0C_+C_-C_{m,\lambda})^2~e^{-i(E-E')T}\int^{+\infty}_{-\infty} e^{-i(K_0-K'_0)Z_0}~dZ_0 \int^{+\infty}_{-\infty} e^{-i(K_+-K'_+)Z_+}~dZ_+ \int^{+\infty}_{-\infty} e^{-i(K_--K'_-)Z_-}~dZ_- \\ \nonumber
\times\int^{\infty}_{0} \kappa ^{\frac{1}{m+2}} {\kappa'} ^{\frac{1}{m+2}} \phi J_{\frac{1}{m+2}}\left(\frac{2\sqrt{\lambda} \phi^{\frac{m+2}{2}} \kappa}{m+2}\right) F(\phi) J_{\frac{1}{m+2}}\left(\frac{2\sqrt{\lambda} \phi^{\frac{m+2}{2}} \kappa'}{m+2}\right)~d\phi \\ \nonumber
=(C_0C_+C_-C_{m,\lambda})^2~e^{-i(E-E')T}\delta(K_0-K'_0)\delta(K_+-K'_+)\delta(K_--K'_-)\int^{\infty}_0 \kappa ^{\frac{1}{m+2}} {\kappa'} ^{\frac{1}{m+2}} \phi J_{\frac{1}{m+2}}\left(\frac{2\sqrt{\lambda} \phi^{\frac{m+2}{2}} \kappa}{m+2}\right) \\ \nonumber
\times~ F(\phi) J_{\frac{1}{m+2}}\left(\frac{2\sqrt{\lambda} \phi^{\frac{m+2}{2}} \kappa'}{m+2}\right)~d\phi~.
\end{gather}
To perform the $\phi$ integral we put the form of $F(\phi)=\frac{\lambda}{4} \phi^m$ and then make a variable change $\frac{2\sqrt{\lambda} \phi^{\frac{m+2}{2}}}{m+2}=Y$. The $\phi$ integration then reduces to the form
\begin{equation}
\label{3.18}
\frac{1}{4}\left(\frac{1}{2\gamma}\right)^2 {\kappa'}^\gamma\kappa^{\gamma-1}\int^\infty_0\kappa Y J_\gamma(Y\kappa)J_\gamma(Y\kappa')~dY
\end{equation}
where $\gamma=\frac{1}{m+2}$. Making use of the result $\int_0^\infty dq~ xq J_\beta (xq)J_\beta (aq)=\delta(x-a)$ to perform the $\phi$ integral gives the value of the full integral to be
\begin{gather}
\label{3.19}
\int dZd\phi~ \Psi(Z,\phi,T) F(\phi) \Psi^{'*}(Z,\phi,T)\\ \nonumber
=(C_0C_+C_-C_{m,\lambda})^2\delta(K_0-K'_0)\delta(K_+-K'_+)\delta(K_--K'_-)\delta(\kappa-\kappa')\left(\frac{1}{4}\left(\frac{1}{2\gamma}\right)^2{\kappa'}^\gamma\kappa^{\gamma-1}\right) e^{-i(E-E')T}~.
\end{gather}
Putting this result into eq.(\ref{3.16}), we get
\begin{eqnarray}
\label{3.20}
||\Psi_{wp}||&=&\frac{1}{4}\left(\frac{C_0C_+C_-C_{m,\lambda}}{2\gamma}\right)^2\int e^{-2(K_0^2 + K_+^2 + K_-^2+\kappa^2)}~ d\kappa dK_0 dK_+ dK_-\\ \nonumber
&=&\frac{1}{4}\left(\frac{C_0C_+C_-C_{m,\lambda}}{2\gamma}\right)^2\frac{1}{2}\left(\sqrt{\frac{\pi}{2}}\right)^4~.
\end{eqnarray}
Note that the norm of the wave packet is independent of time.
So now the normalized wave packet becomes
\begin{equation}
\label{3.20a}
\Psi_{wp}=\frac{8\sqrt{2}\gamma}{\pi C_0C_+C_-C_{m,\lambda}}\int \kappa^{\frac{1}{2}-\gamma}e^{-(K_0^2 + K_+^2 + K_-^2+\kappa^2)}\Psi(Z,\phi,T)~d\kappa dK_0 dK_+ dK_-~.
\end{equation}
From this wave packet (\ref{3.20a}), one can now proceed to calculate the expectation value of the spatial volume of the universe. This reads
\begin{gather}
\label{3.21}
\langle ABC \rangle (T) \equiv \langle e^{3Z_0} \rangle(T) \\ \nonumber
=\bigg|\frac{8\sqrt{2}\gamma}{\pi C_0C_+C_-C_{m,\lambda}}\bigg|^2\int dZ d\phi \int \kappa^{\frac{1}{2}-\gamma}e^{-(K_0^2 + K_+^2 + K_-^2+\kappa^2)}\Psi(Z,\phi,T)~d\kappa dK_0 dK_+ dK_- \\ \nonumber
\times~e^{3Z_0(T)}F(\phi)\int \kappa'^{\frac{1}{2}-\gamma}e^{-(K_0^{'2} + K_+^{'2} + K_-^{'2}+{\kappa'}^2)}\Psi^{'*}(Z,\phi,T)~d\kappa' dK'_0 dK'_+ dK'_- ~.
\end{gather}
Evaluating the above integrals we get the result
\begin{gather}
\label{3.22}
\langle e^{3Z_0} \rangle(T)=2\left(\sqrt{\frac{2}{\pi}}\right)^4\int \Big[ e^{-K_0^2}e^{-iK_0 Z_0} e^{3Z_0} e^{-K_0^{'2}}e^{iK'_0 Z_0}~dZ_0 \nonumber\\ 
\times\int e^{-2(K_+^2 + K_-^2+\kappa^2)}e^{-i(E-E')T}~ d\kappa dK_+ dK_-\Big]dK_0dK'_0 \nonumber\\ 
=\sqrt{\frac{2}{\pi}}\int e^{-K_0^2}e^{-iK_0 Z_0} e^{iK_0^2T} e^{3Z_0} e^{-K_0^{'2}}e^{iK'_0 Z_0}e^{-iK_0^{'2}T}~dZ_0dK_0dK'_0 \nonumber\\
=\sqrt{\frac{2}{\pi}}\int \frac{1}{2\sqrt{(1+T^2)}}e^{-\frac{Z_0^2}{2(1+T^2)}}e^{3Z_0}~dZ_0 \nonumber\\ 
=e^{\frac{9}{2} \left(T^2+1\right)}~~~; \hskip 2cm\Re\left(T^2\right)>-1~.
\end{gather} 
It clearly tells us that at the beginning of time, that is at $T=0$, the universe had a finite volume. Figure 1 displays the variation of the volume expectation of the universe with the time parameter $T$.
\begin{figure}[h!]
\centering
\begin{subfigure}[b]{0.5\linewidth}
\includegraphics[width=\linewidth]{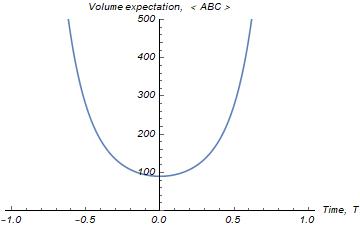}
\end{subfigure}
\caption{\textit{Expectation value of the volume of universe vs. the time parameter T}}
\end{figure}
\noindent Now we proceed to study the behaviour of the probability density function, that is
\begin{gather}
\label{3.25}
\rho = \Psi^*_{wp}\Psi_{wp}\\ \nonumber
=\bigg|\frac{8\sqrt{2}\gamma}{\pi C_0C_+C_-C_{m,\lambda}}\bigg|^2\int \kappa^{\frac{1}{2}-\gamma}e^{-(K_0^2 + K_+^2 + K_-^2+\kappa^2)}\Psi(Z,\phi,T)~d\kappa dK_0 dK_+ dK_- \\ \nonumber
\int \kappa'^{\frac{1}{2}-\gamma}e^{-(K_0^{'2} + K_+^{'2} + K_-^{'2}+{\kappa'}^2)}\Psi^{'*}(Z,\phi,T)~d\kappa' dK'_0 dK'_+ dK'_- ~.
\end{gather}
Using the form of $\Psi$ from eq.(\ref{3.14}) and also using eq.(\ref{3.10a}), we obtain the expression for the probability density function 
\begin{gather}
\label{3.26}
\rho=\left(\frac{8\sqrt{2}\gamma}{\pi}\right)^2\frac{1}{8(1+T^2)^{3/2}}e^{\frac{-(Z_0^2+Z_+^2+Z_-^2)}{2(1+T^2)}} \phi \int^\infty_0 \sqrt{\kappa}e^{-\kappa^2} e^{i\kappa^2 T} J_{\frac{1}{m+2}}\left(\frac{2 \phi^{\frac{m+2}{2}} \kappa}{m+2}\right) d\kappa \\ \nonumber
\int^\infty_0 \sqrt{\kappa'}e^{-\kappa'^2}e^{-i\kappa'^2 T} J_{\frac{1}{m+2}}\left(\frac{2 \phi^{\frac{m+2}{2}} \kappa'}{m+2}\right) d\kappa'\\ \nonumber
=\frac{\lambda}{16\pi^2}\gamma^{2(\gamma+1)}\Gamma\left(\frac{\gamma}{2}+\frac{3}{4}\right)^2(1+T^2)^{-\frac{\gamma}{2}-\frac{9}{4}}\phi^2e^{\frac{-(Z_0^2+Z_+^2+Z_-^2)}{2(1+T^2)}} ~_1\tilde{F}_1\left(\frac{\gamma }{2}+\frac{3}{4};\gamma +1;-\frac{\gamma ^2 \lambda  \phi
	^{\frac{1}{\gamma }}}{i T+1}\right) \, \\ \nonumber
\times~ _1\tilde{F}_1\left(\frac{\gamma }{2}+\frac{3}{4};\gamma +1;\frac{\gamma ^2 \lambda  \phi ^{\frac{1}{\gamma }}}{i T-1}\right)
\end{gather} 
where $_1\tilde{F}_1$ is the regularized hypergeometric function. Fig.2 displays the variation of the probability density function with respect to $\phi$ and $Z_0$ considering $Z_+=Z_-=0$ for three different values of the time parameter $T$. It can be seen that the height of the peak reduces with increase in the time  parameter $T$
.
\begin{figure}[h!]
\centering
\begin{subfigure}[b]{0.45\linewidth}
\includegraphics[width=\linewidth]{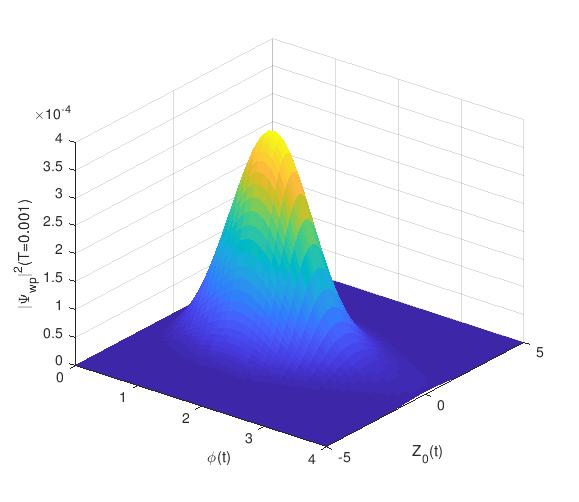}
\caption{$T=0.001$}
\end{subfigure}
\begin{subfigure}[b]{0.45\linewidth}
\includegraphics[width=\linewidth]{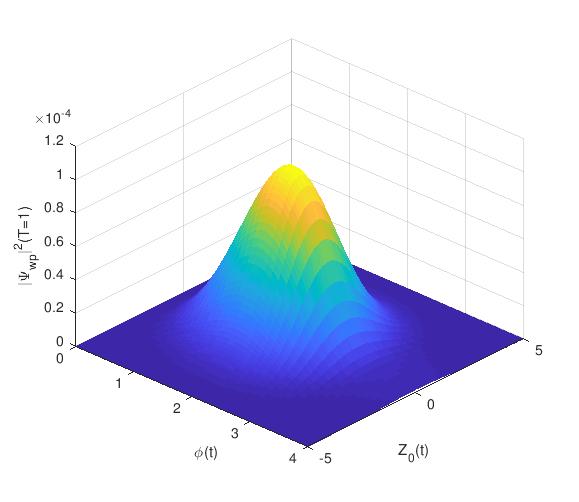}
\caption{$T=1$}
\end{subfigure}
\begin{subfigure}[b]{0.45\linewidth}
\includegraphics[width=\linewidth]{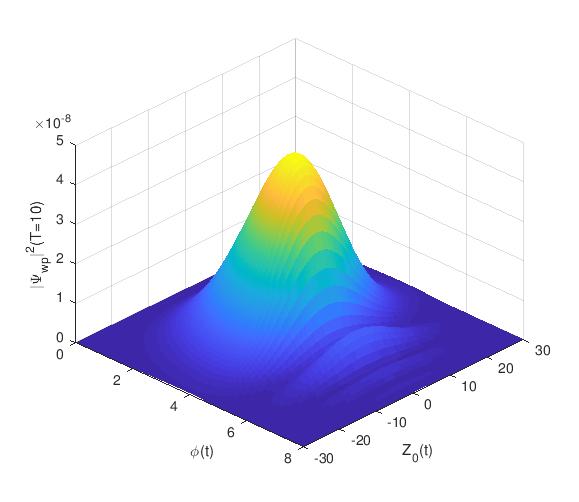}
\caption{$T=10$}
\end{subfigure}
\caption{\textit{Behavior of the probability density function with respect to $\phi$ and $Z_0$. We plot for three different values of the time parameter $T$, with the values $\lambda=1$ and $m=2$.}}
\label{Fig.1}
\end{figure}

\section{Conclusions}
In this paper we have explored the quantum cosmology of a scalar-metric gravity theory in the presence of a perfect fluid as the cosmic matter. We have also considered an anisotropic cosmological model, namely, the Bianchi type $I$ metric which is the simplest among all other anisotropic models. We make use of the well known Schutz formalism to deal with the matter sector. The matter sector variable plays the role of the time parameter. The full theory is then quantized by the Wheeler-DeWitt prescription. 
To have a consistent theoretical framework, we define a proper inner product between two wave functions that makes the quantum Hamiltonian a self-adjoint operator. With proper boundary conditions we then construct the wave packet for the universe. This wave packet gives an explanation for the singularity free birth of the universe. In the entire analysis we have considered a particular value for the equation of state parameter that is $\omega=1$. The generalization can be easily made for $\omega\neq1$. Of course for $\omega\neq1$ one should seriously consider the operator ordering issue.

From the structure of the wave function obtained, we observe that the quantum mechanical evolution of the universe is nonunitary. This is a feature which has been observed in the literature for anisotropic cosmological models and appears in our analysis also. In this regard we would like to point out that such a nonunitary quantum evolution of the universe was also seen in 
\cite{Saumya et. al.} for FRW metric which is an isotropic cosmological model. This actually owed its origin to the inclusion of the scalar field in the model. In this study, we restore the unitarity by a suitable choice of the weight factor in the construction of the wave packet rendering the norm of the wave packet time independent.


\noindent We also analyze the behaviour of the wave packet that we construct. Firstly, the expectation value of the volume shows an inflationary expansion with respect to the time parameter.
This is displayed in Figure 1.
Figure $2$ depicts the nature of the probability density function of the wave packet. We have analyzed it for three different values of the time parameter $T$. At very low value of $T(=0.001)$, we see that the peak of the density function has the highest value. As time increases the function spreads over the $Z_0$ and $\phi$ domain. 
\section*{Acknowledgment}
S.G. acknowledges the support by DST-SERB under Start Up Research Grant (Young Scientist), File No.YSS/2014/000180. He also acknowledges the support of IUCAA, Pune for the Visiting Associateship.  S. Ghosh would like to thank Arnab Acharya of IISER-Kolkata, for his help in preparing this manuscript. 

\section*{Author contribution statement}
All authors have contributed equally.


\end{document}